\documentclass[fleqn,usenatbib]{mnras}

\usepackage{newtxtext,newtxmath}

\usepackage[T1]{fontenc}
\usepackage{ae,aecompl}

\usepackage{graphicx}	
\usepackage{amsmath}	
\usepackage{amssymb}	

\title[high z galaxies]{Inception of  a first quasar at cosmic dawn}

\author[Latif  et al.]{
Muhammad A. Latif,$^{1}$\thanks{E-mail: latifne@gemail.com} 
Sadegh Khochfar$^{2}$  \\
$^{1}$Physics Department, College of Science, United Arab Emirates University, PO Box 15551, Al-Ain, UAE \\
$^{2}$Institute for Astronomy, University of Edinburgh, Royal Observatory, Blackford Hill, Edinburgh EH9 3HJ, UK }

\date{Accepted XXX. Received YYY; in original form ZZZ}

\pubyear{20202}

\begin{document}
\bibliographystyle{mnras}

\label{firstpage}
\pagerange{\pageref{firstpage}--\pageref{lastpage}}
\maketitle

\begin{abstract}
Earliest quasars at the cosmic dawn are powered by mass accretion onto supermassive black holes of a billion solar masses. Massive black hole seeds  forming through the direct collapse mechanism are considered the most promising candidates but how do they grow and co-evolve with their host galaxies at early cosmic times remains unknown.  We here present results from  a cosmological radiation hydrodynamical simulation including self-consistent  modeling of both Pop III and Pop II star formation, their  radiative and supernova feedback in the host galaxy along with X-ray feedback from an accreting massive black hole (MBH) of $\rm 10^5 ~M_{\odot}$ in  a halo  of $\rm 2 \times 10^9~M_{\odot}$ from $z=26$ down to $z=16$.  Our results show that energy deposition from X-rays in the  proximity of MBH suppresses Pop III star formation for about 12 Myr while at the same time these X-rays catalyze $\rm H_2$ formation which leads to the formation of a Pop III  star cluster of 500 $\rm M_{\odot}$  in the close vicinity of the MBH.  We find that mode of star formation for Pop III  is  episodic and bursty  due to  the clumpy accretion while for  Pop II it is continuous. The stellar mass of the host galaxy at $z \sim 16$ is $\rm 2 \times 10^7~M_{\odot}$ with a star formation rate (SFR) of $\rm \sim 0.1-1~M_{\odot}/yr$. In total, the MBH  accretes $\rm 1.5 \times 10^6~M_{\odot}$ during 120 Myr with the mean accretion rate of $\rm \sim 0.01~M_{\odot}/yr$ corresponding to an average Eddington fraction of 50\%.   

\end{abstract}
\begin{keywords}
methods: numerical -- cosmology: theory -- early Universe -- high redshift-- galaxies formation-- black holes
\end{keywords}



\section{Introduction}

Most present-day galaxies  harbour  supermassive black holes (SMBHs) of a few times $\rm 10^6 -10^9 ~M_{\odot}$ at their centers. The correlation between  their  bulge component and the central  SMBH  the so-called $M-\sigma$ relation  suggests  their co-evolution \citep{Silk98,Kormendy13}.  On the other hand, observations of quasars  reveal SMBHs of   $\rm \sim 10^9 ~M_{\odot}$ at $z>6$ \citep{Fan2003,Jiang2009, Mortlock2011,Venemans2015,Banados18,Matsuoka19}. Seeds of these SMBHs are  expected to be formed  at $z=20-30$ \citep{Latif16PASA} which comprise remnant  BHs from Pop III stars \citep{Heam20},  BHs forming  from the core collapse of dense stellar clusters through stellar dynamical processes \citep{Yajima16}  and massive BHs resulting from monolithic collapse of  protogalactic gas clouds  known as the direct collapse black holes (DCBHs) \citep{Agarwal14,Agarwal2015B}. For a detailed description and discussion of these mechanisms, see dedicated reviews in \citet{Volonteri2010,Latif16PASA,Woods19,Inayoshi19}.  Concurrently, the first galaxies form  in  dark matter halos of a few times $\rm 10^7~M_{\odot}$ massive enough to host stellar populations a few hundred years after the Big Bang.  In fact, this picutre is confirmed by high redshift galaxy surveys which have detected more than 800 galaxies above $z>6$ with candidates up to $z=11$ \citep{Oesch16,Lam19,Bowler20}. However, how the earliest MBHs  grow and co-evolve with their host galaxies at cosmic dawn still remains unknown.

Primordial stars, so-called Pop III stars, form in  pristine minihalos of $\rm 10^5-10^6~M_{\odot}$ at $z=20-30$ predominantly cooled by molecular hydrogen formed out of gas phase reactions and ushered the cosmos out of cosmic dark ages \citep[e.g.][]{Bromm2013}. While pioneering studies proposed that Pop III stars are more massive with typical masses of a few hundred solar \citep{Abel2000,Bromm2002,Yoshida2003} this paradigm has been challenged  during the past decade. Recent high resolution simulations including feedback from protostars advocate for the multiplicity of  Pop III stars and suggest  characteristic masses of a few tens of solar \citep{Clark11,Greif12,Latif13ApJ,Stacy16,Hajima19,Sugimura20}. Depending on the mass spectrum, these Pop III stars regulate star formation in their host galaxies through radiative, chemical and mechanical feedback \citep{Whalen04,Whalen2013b,Latif19}. They exploded into supernovae (SNe) and enriched the universe with metals \citep{Heger2002} and lead to the formation of  the second generation of stars known as Pop II stars. The latter typically form from gas phases with metallicity  above a critical metallicity of  $\rm Z/Z_{\odot} = 3 \times 10^{-4}$ \citep{Schneider03,Omukai05,Hartwig19}. Eventually both stellar populations are expected to reside in the first galaxies \citep{Maio11,Johnson13}. Some of the Pop III stars may form in binaries, leading to X-ray binaries which may  influence subsequent star formation \citep{Jeon14,Ryu16} .

Pop III stars of  masses between $\rm 25-140~M_{\odot}$ and above $\rm 260~M_{\odot}$ are expected to collapse into  BHs \citep{Heger2002,Heger2003}. These stellar mass BHs  are born in HII/low density regions  and are therefore subject to ejection by natal kicks  which leads to their stunted growth \citep{Johnson2007,Alvarez09,Milos2009,Park2011,Jeon12,Whalen12,Smith18}.  The other channel for BH  formation could be dynamical  evolution of a dense star cluster. In this case runaway stellar collisions may get triggered in the first dense stellar clusters forming at $z \sim10-15$ during the core collapse of such cluster and may result in a very massive star which subsequently collapses into a black hole of up to a thousand solar masses \citep{Omukai2008,Devecchi2012,Yajima16,Latif2016dust,Chon20}. Recent studies suggest that collision may further enhance BH masses upto $\rm 10^5~M_{\odot}$ \citep{Reinoso18a,Tagawa20,Woods20}. Alternatively, a promising mechanism is  the direct collapse scenario which provides  a massive BH (MBH) of $\rm \sim 10^5~M_{\odot}$ about two orders of magnitude more massive than other mechanisms. The prerequisite  for this scenario is large inflow rates of $\rm 0.1 ~M_{\odot}/yr$ which can be obtained more easily  thermodynamically under isothermal conditions in metal free halos  \citep{Agarwal2012,Latif2013d, Shlosman2016,Regan18b,Becerra18,Chon18,Agarwal19, Latif20} or alternatively dynamically via gas rich galaxy mergers \citep{Mayer15}.

Large scale cosmological simulations investigating the growth of MBH generally employ thermal feedback without  proper radiative transfer \citep{Booth209,Dubois15,Sijacki15,Alcazar17,DiMatt17}. They have typical resolution of  a few hundred parsec, dark matter (DM)  particle  resolution of $\rm \sim 10^8~M_{\odot}$ and are therefore unable to study the growth of MBH with their  host galaxies at  $z \geq 10$. Moreover,  they  may overestimate the mass accretion onto MBHs by not resolving the Bondi radius \citep[see e.g.][for a discussion]{Gaspari2013,Negri2017} for discussion.
Radiative transfer cosmological simulations exploring the growth of MBH in halos  of $\rm 10^7~M_{\odot}$  \citep{Johnson2011} and  $\rm 10^8~M_{\odot}$ \citep{Aykutalp13,Aykutalp14} find that mass accretion rate onto MBH varies from $\rm 10^{-7}-10^{-3}~M_{\odot}/yr$ corresponding to duty cycle of 50 \% fraction and X-ray feedback stifles its growth. \cite{Smidt17} performed radiation hydrodynamical simulations and found that cold accretion flows feed  MBHs which consequently grow to billion solar masses by $z=7$. The DM resolution in their simulation is $\rm 8.4 \times 10^{6}~M_{\odot}$ and the minimum stellar particle mass is $\rm 10^7 ~M_{\odot}$, which did not allow them to resolve atomic cooling halos and pop III star formation in their simulation. \cite{Latif18}  explored the growth of a  MBH in $\rm 3 \times 10^{10}~M_{\odot}$ halo at $z=7.5$ including both UV and X-ray feedback from MBHs as well as chemical, mechanical and radiative feedback from both Pop III and Pop II stars. They found that feedback from the MBH in combination with supernova feedback  expel the gas from its vicinity, shutoff gas accretion which results in stunted MBH growth. However, they turned on star formation and feedback from the MBH at $z=12$ in a halo of $\rm 2 \times 10^9~M_{\odot}$ missing prior episodes of star formation and feedback impact on the evolution of the MBH and host galaxy. They also ignored the cooling from metals produced from Pop II stars. None of these simulations have been able to study the growth of MBHs along with star formation in the first galaxies at $z>15$. 
Depending on their IMF Pop III stars may explode as pair-instability SNe which are 10-100 times more energetic than type II SNe and therefore more effective in removing the gas from halos. Also  Pop III stars produce more high energy radiation than Pop II stars and create HII regions which leave behind a low density medium. Similarly  PISNe from Pop III stars produce higher metal yields which may lead to more cooling and star formation. These processes can strongly influence the growth of MBH and thus self-consistent modelling the feedback from Pop III stars is absolutely necessary. Also X-rays from MBH may regulate star formation by heating the gas and catalyzing the formation of molecules. Such interplay between stellar and MBH feedback sets the stage for its co-evolution with host galaxies which remains poorly known.

In this work  we investigate the growth of a DCBH in halo of $\rm 10^9~M_{\odot}$ and study its co-evolution with its host galaxy  from $z= 26$ down to $z=16$ corresponding to an time scale of 120 Myr. We perform a cosmological radiation hydrodynamical simulation by self-consistently modeling chemical, mechanical and radiative feedback from both Pop III and Pop II stars along with X-ray feedback from an accreting DCBH of $\rm 10^5~M_{\odot}$ forming in the most progenitor halo of the host galaxy at $z=26$. We simulate the radiative feedback from each star particle and X-ray feedback from a MBH as well as every stellar mass BH. At $z=16$, we have about several $10^3$  radiation sources for which we compute the radiative transfer on the fly. We also self-consistently model the transition from Pop III to Pop II stars by simulating both PISNe and core collapse SNe.
Modeling of detailed physical processes and high resolution of a few pc enables us to robustly study the growth of a MBH along with the co-evolution of the first galaxy at such earlier times.  This simulation also bridges the gap between large scale cosmological simulations (exploring the growth of MBHs in $\rm 10^{10}~M_{\odot}$ halos at $z< 10$ using thermal feedback) and small scale radiation cosmological simulation in halos of $\rm \leq 10^8 ~M_{\odot}$.  We discuss  recipes of star formation, stellar feedback and accretion onto the MBH in section 2. The growth of the MBH, star formation and co-evolution  in the host galaxy are  examined in section 3. We present our conclusion in section 4.

\section{Numerical methods}

We  conduct  a cosmological hydrodynamical  simulation coupled with the radiative transfer module MORAY \citep{Wise11} to model   X-ray feedback from the accreting MBH as well as stellar mass black holes  and radiative feedback from each star particle (both Pop III \& Pop II)  using the adaptive mesh refinement code Enzo \citep{Enzo14}. Our simulation uses  cosmological initial conditions generated from the MUSIC package \citep{Hahn2011} at $z =200$ and we use cosmological parameters based on the PLANCK 2016 data with  $\Omega_{\Lambda}=0.6911$, $\Omega_{\rm M}=0.3089$,  $\rm H_{0}=0.6774$ \citep{Planck2016}. The simulated volume has a comoving  size of $\rm 25 ~Mpc/h$  with a top grid resolution of $\rm 256^3$ and three additional nested refinement levels each with resolution of  $\rm 256^3$ grids yielding an effective resolution of $\rm 2048^3$.  We further employ 10 additional refinement levels during the course of simulation which yields in physical resolution of  about 4 pc  and an effective DM resolution of $\rm \sim 10^5 ~M_{\odot}$.  Our refinement criteria is based on baryonic over-density, particle mass resolution and the Jeans refinement  of  at least four cells, similar to \cite{Latif18}.

Our simulated halo has a mass of  $\rm 2 \times 10^9~M_{\odot}$ at $z \sim 16$ and is placed at the centre of the computational box. We turn on X-ray feedback from the MBH of $\rm 10^5~M_{\odot}$ assumed to have formed by the direct collapse mechanism \citep{Latif20}  in the most massive progenitor of the halo which reaches the atomic cooling threshold at $z=26$ and we simultaneously allow star formation along with radiative, chemical and mechanical feedback. The radiation transport module MORAY is coupled to  a non-equilibrium primordial chemistry solver which solves the rate equations of the following species $ \rm H, ~H^+,~ H^-, ~He,~ He^+, ~He^{++},~ H_2, ~H_2^+, ~e^-$ \citep{Abel97}.  We assume a background UV flux of strength 500 in units of $\rm J_{21}=10^{-21}~erg/cm^2/s/Hz/Sr$. Our chemical model includes Compton heating/cooling, $\rm H_2$ cooling,  cooling due to the collisional excitation and  collisional ionization, Bremsstrahlung radiation and  radiative recombination. We also include metallicity dependent  metal line cooling  (C, O, N, Si etc) from \cite{Glover07} which operates in $\rm 100-10^4~K$ regime and for temperatures $\geq 10^4$ K we employ tabulated cooling functions from \cite{Sutherland93}.  We also model  photoionization heating and secondary electron ionization heating \citep{Shull85} from X-rays emitted by MBH. 

\subsection {Star formation and stellar feedback}
Our recipes for star formation (both Pop III \& Pop II stars) and feedback are based on \cite{Wise08A}  \& \cite{Wise12} and  same as in  \cite{Latif18}. We present here a brief summary and for details  refer reader to the above mentioned references.   A Pop III star particle is created in a cell meeting the following three criteria, 1)  an over density of $5 \times 10^5$ ($\rm 10^3 ~cm^{-3}$ at $z=10$, II) a molecular hydrogen ($\rm H_2$) fraction of $ \rm \geq 5 \times 10^{-4}$, III) convergent flow ($\rm \nabla  \cdot v_{gas} <0 $).  We discern between Pop III and Pop II stars based on metallicity,  for $\rm Z/Z_{\odot} \leq 10^{-4}$ Pop III stars are formed otherwise Pop II stars \citep{Schneider03,Omukai2008}. In our simulation each Pop III star particle represents a single star and its mass is randomly drawn from the initial mass function (IMF) with mass range between 1-300 $\rm M_{\odot}$ which behaves like the Salpeter IMF above cut off mass of $\rm 100~M_{\odot}$ and an exponential shape below it. In our simulation a Pop II star particle represents a small cluster of stars and  their formation criteria is  the same as for Pop III  except  that the condition of molecular hydrogen fraction is removed.

Star particles are treated as point sources and the feedback from them is modeled using the  adaptive ray tracing algorithm  based on the HEALPix  scheme \citep{AbelWand02,Wise11} coupled with hydrodynamics. We consider Pop III and Pop II stars as monochromatic sources with energy of 29.6 eV and 21.6 eV, respectively.  The mass dependent luminosities for Pop III stars are taken from \cite{Schaerer02} while  Pop II stars produce 6000 photons per stellar baryon for 20 Myr (equivalently $\rm 2.4 \times 10^{47} photons/s/M_{\odot}$, \cite{Schaerer2003}). Pop III stars either die as SNe or collapse into BHs depending on their mass \citep{Heger2002,Heger2003}. We model both type II SNe and pair instability SNe  from Pop III stars as well as X-ray feedback from each BH particle using MORAY. Pop II stars produce $\rm 6.8 \times 10^{48} erg/s/M_{\odot}$ from SNe 4 Myr after their formation which is distributed in a sphere of 10 pc radius \citep{Woosley86,Wise12}.

\subsection {MBH  accretion and feedback}
We insert a MBH of $\rm 10^5~M_{\odot}$ assumed to have formed via the DC scenario at the center of the first atomic cooling progenitor halo which appears at $z=26$. The MBH is treated as a sink particle and grows via mass accretion.  The mass accretion onto the MBH is estimated using  the Eddington limited Bondi Hoyle formalism \citep{Bondi44,Bondi52} \citep[see][for a detailed prescription]{Kim11}. We model the luminosity of an accreting MBH as $L_{\rm{MBH}}= \epsilon_{r} \dot{M}_{\rm{BH}} c^2$ where $\rm \epsilon_{r}$ is the radiative efficiency assumed to be 0.1 \citep{Shakura73},  $\dot{M}_{BH}$  is the mass accretion rate onto the  MBH and $c$ is the speed of light.  We assume here that all the luminosity from the MBH is emitted at 2 KeV, corresponding to the averaged quasar spectral energy distribution \citep{Sazonov04,Ciotti17} and is also consistent with observations of quasars suggesting 90\% of their X-ray flux comes from 0.5-2 keV \citep{Nanni17}. The Bondi radius for gas with temperature of 8000 K is about 8.6 pc which is resolved  most of the times  in our simulation.  The radiative feedback from the MBH is modeled  with the 3D radiative transfer module MORAY coupled to the hydrodynamics.

\begin{figure*}
\hspace{-6.0cm}
\centering
\begin{tabular}{c c}
\begin{minipage}{6cm}
\vspace{-0.2cm}
\includegraphics[scale=0.5]{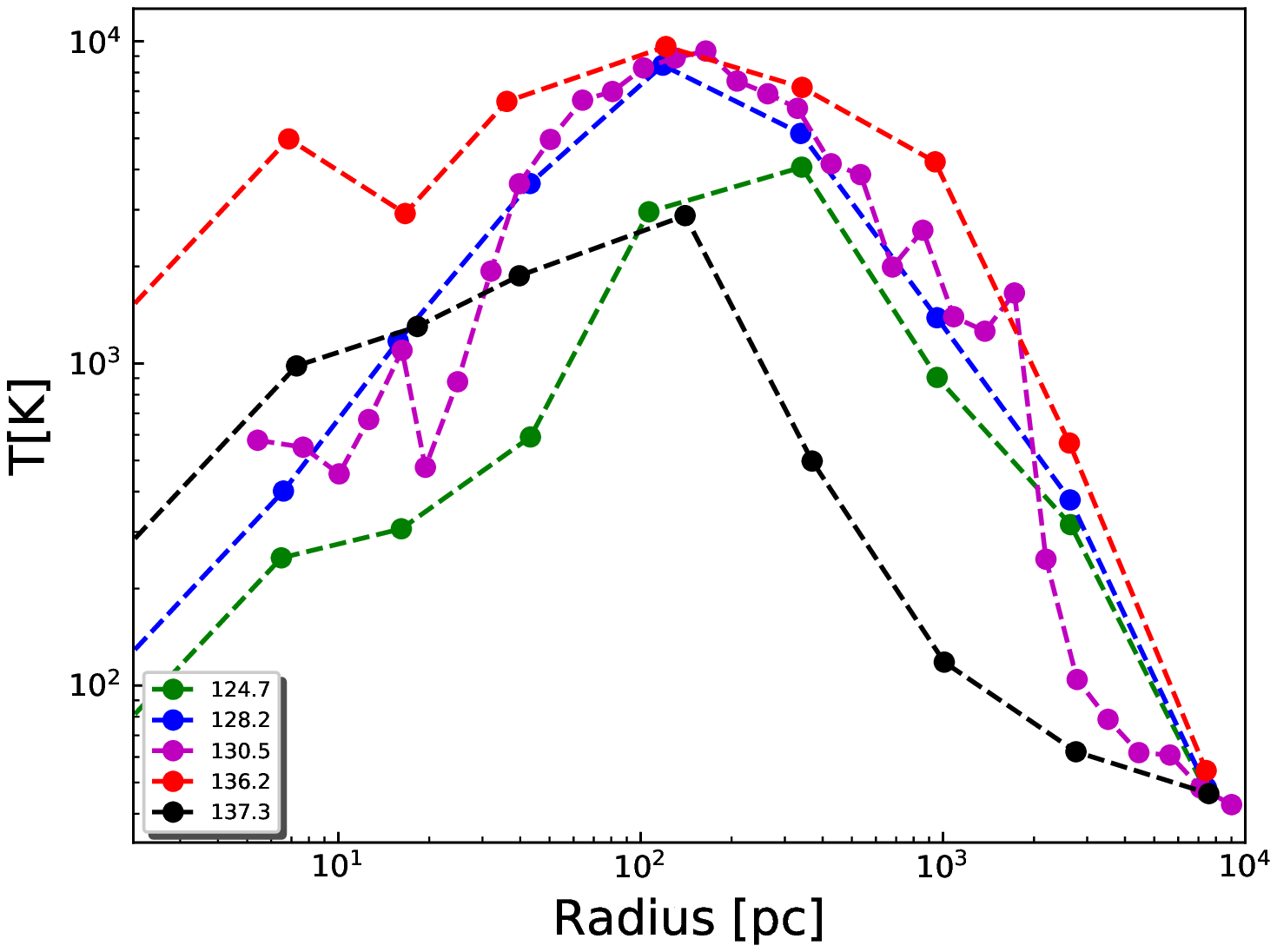}
\end{minipage} &
\begin{minipage}{6cm}
\hspace{1.8cm}
\includegraphics[scale=0.5]{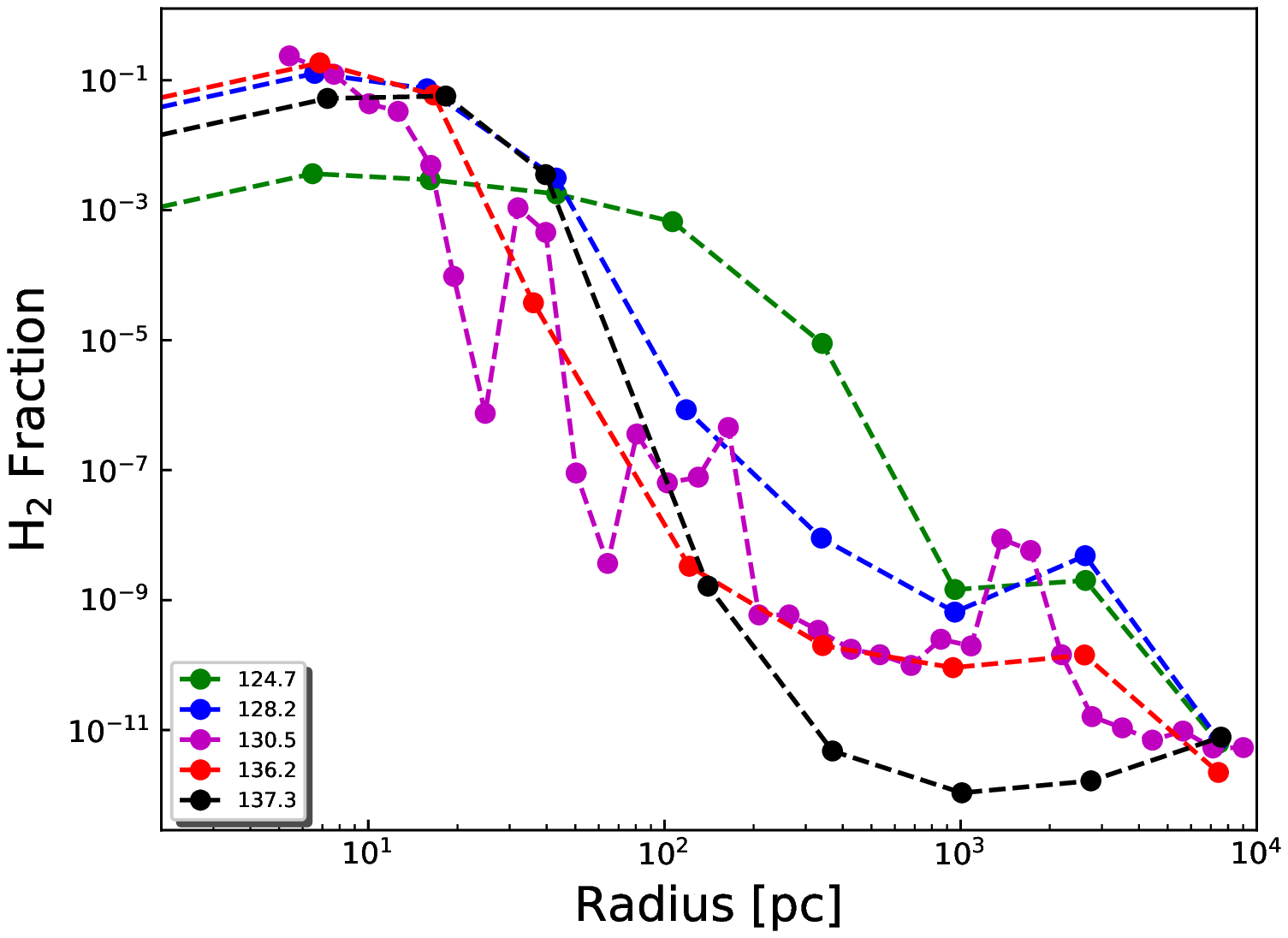}
\end{minipage} \\
\begin{minipage} {6cm}
\vspace{-0.2cm}
\includegraphics[scale=0.5]{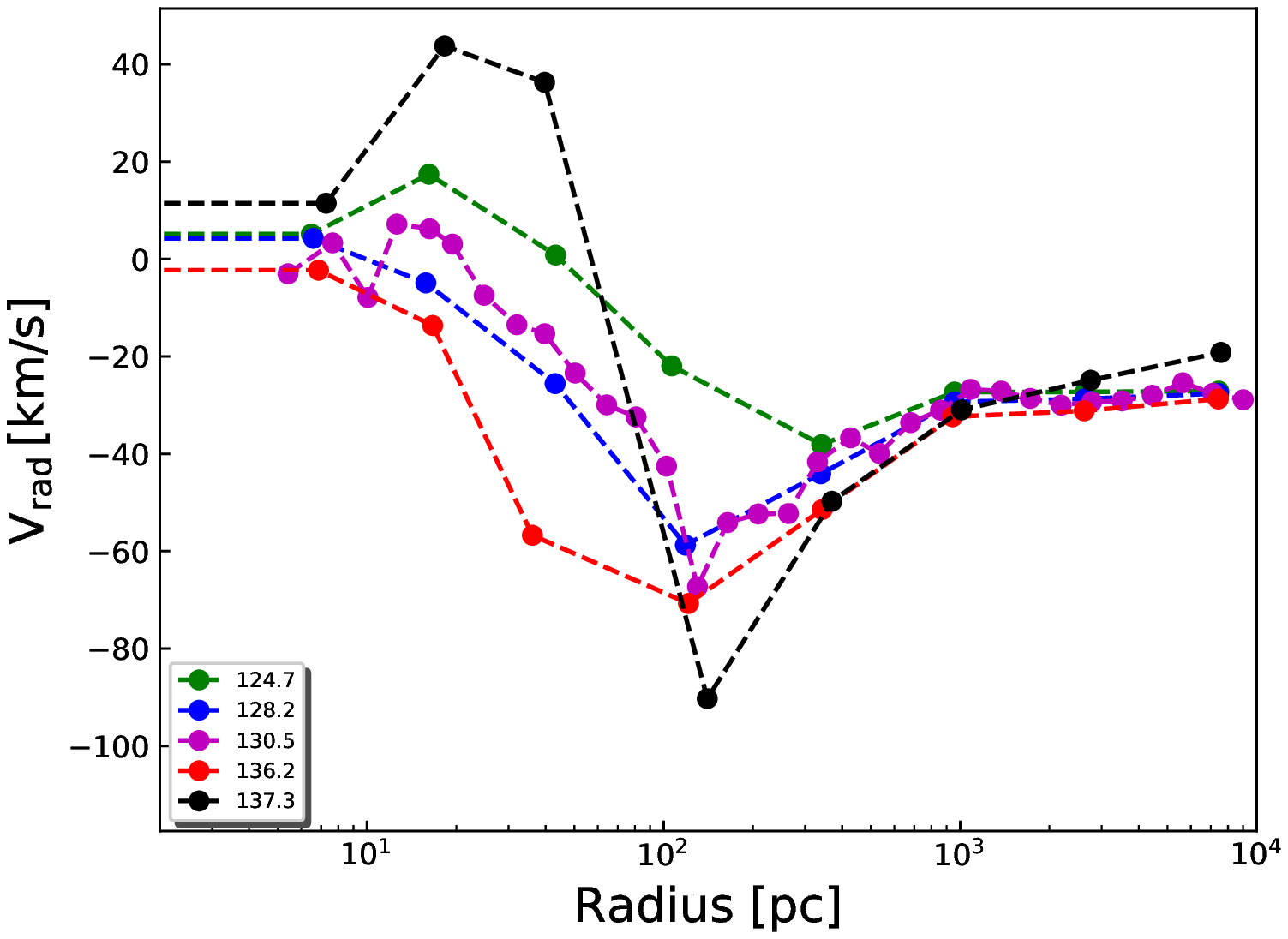}
\end{minipage} &
\begin{minipage}{6cm}
\hspace{1.8cm}
\includegraphics[scale=0.5]{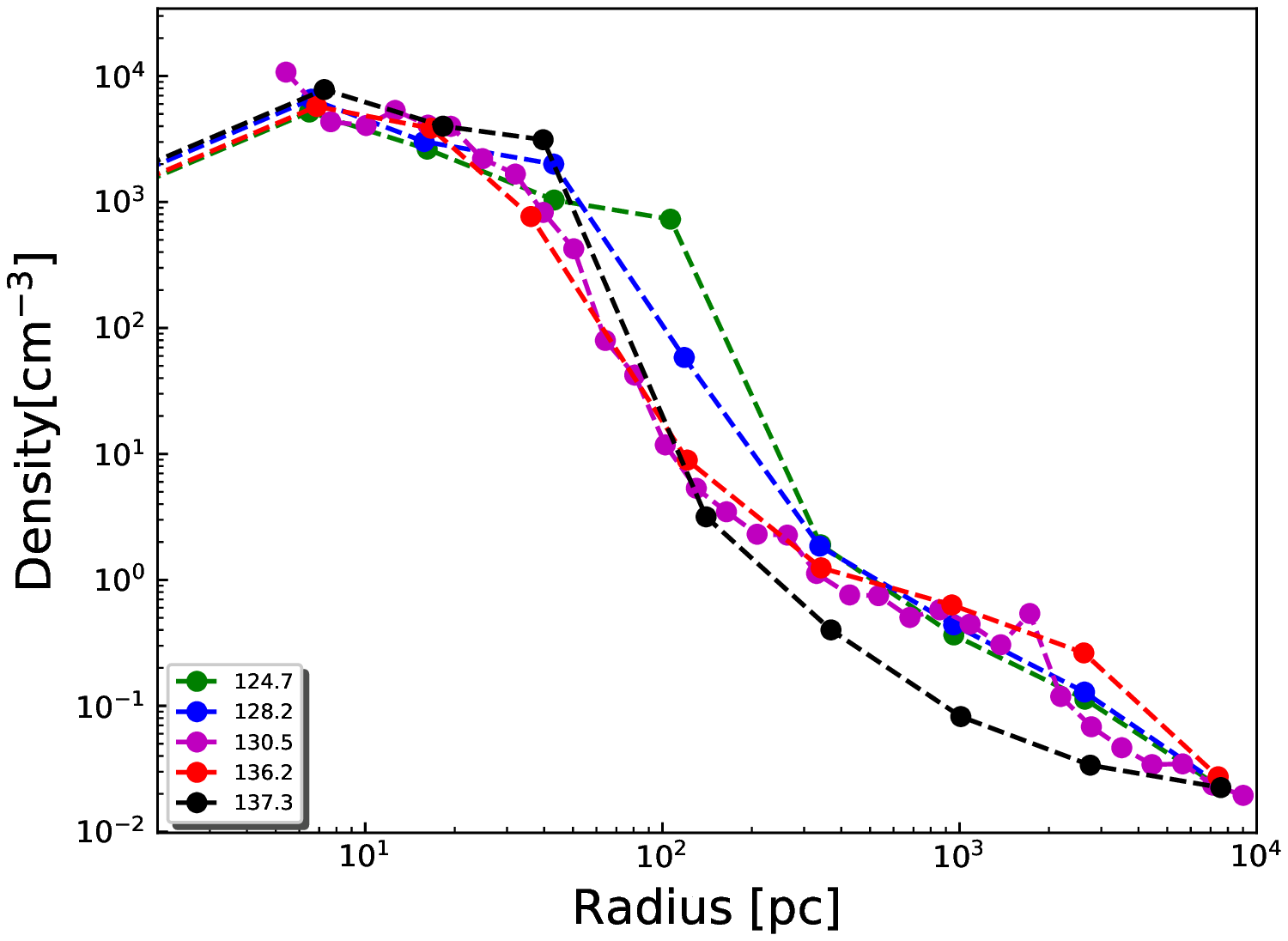}
\end{minipage}
\end{tabular}
\caption{Time evolution of radial velocity,   $\rm H_2$ fraction,  gas density and temperature  prior to  the pop III star formation in host galaxy are shown in this figure.}
\label{fig4}
\end{figure*}

\section{Results}

\subsection{Star formation in the host galaxy}
We seed the MBH at $z \sim 26$, and simultaneously turn on  star formation including chemical, mechanical and radiative feedback along with  X-ray emission from the MBH.  The first Pop III star forms after 12 Myr. This delay in star formation (SF) is caused by the X-ray energy deposition from the MBH which heats the gas and reverses the gas inflow in the host halo. At the same time X-rays catalyze $\rm H_2$ formation and boost its abundance by two orders of magnitude in the central 30 pc of the halo, see Fig. \ref{fig4}.  Ongoing gas infall results in the formation of dense gas clumps, which are shielded  from X-ray heating and lead to the formation of the first Pop III star of $\rm 85~M_{\odot}$. 
A starburst occurs within  2 Myr after the formation of the first star  and  a small Pop III stellar  cluster of  $\rm 500~M_{\odot}$ forms in the vicinity of MBH.
The most massive Pop III star goes off  SN  after 3 Myr and enriches the gas in its vicinity. The metal cooling  reduces  the local Jeans mass and the first Pop II stellar cluster of  $\rm 300 ~M_{\odot}$ forms.  Subsequently, all stars in the first Pop III stellar cluster die  4 Myr after their formation, they  further enrich the medium with metals leading to the formation of many Pop II stellar clusters and  consequently the stellar mass sharply rises above $\rm 10^6~M_{\odot}$. Gas metallicity in the surrounding of the MBH is $\rm Z/Z_{\odot}\sim 0.01$ well above the critical value for Pop II formation and  Pop III stars stop forming in the vicinity of the MBH between $\rm 153-176$ Myr after the Big Bang (30 Myr after MBH seeding) until the fresh supply  of metal free  gas is brought in by a  merger with a pristine mini-halo.  Consequently, another starburst occurs resulting in a  Pop III star cluster of $\rm 10^3~M_{\odot}$  consisting of massive stars which die within a a few Myr.  Pop III SF get halted again for about 20 Myr due to the lack of metal free gas and efficient metal mixing within the central 500 pc of galaxy. Thereafter merging of dense metal free/poor clumps at 200 Myr and 230 Myr  after Big Bang  produce starbursts resulting in Pop III stellar clusters of about thousand  solar masses. Overall, the Pop III mode of SF is highly episodic and bursty due to the recurring clumpy accretion of metal poor gas onto the main halo. At the end of our simulation the Pop III stellar mass is only $\rm 129~M_{\odot}$.

\begin{figure*}
\hspace{-6.0cm}
\centering
\begin{tabular}{c c}
\begin{minipage}{6cm}
\vspace{-0.2cm}
\includegraphics[scale=0.54]{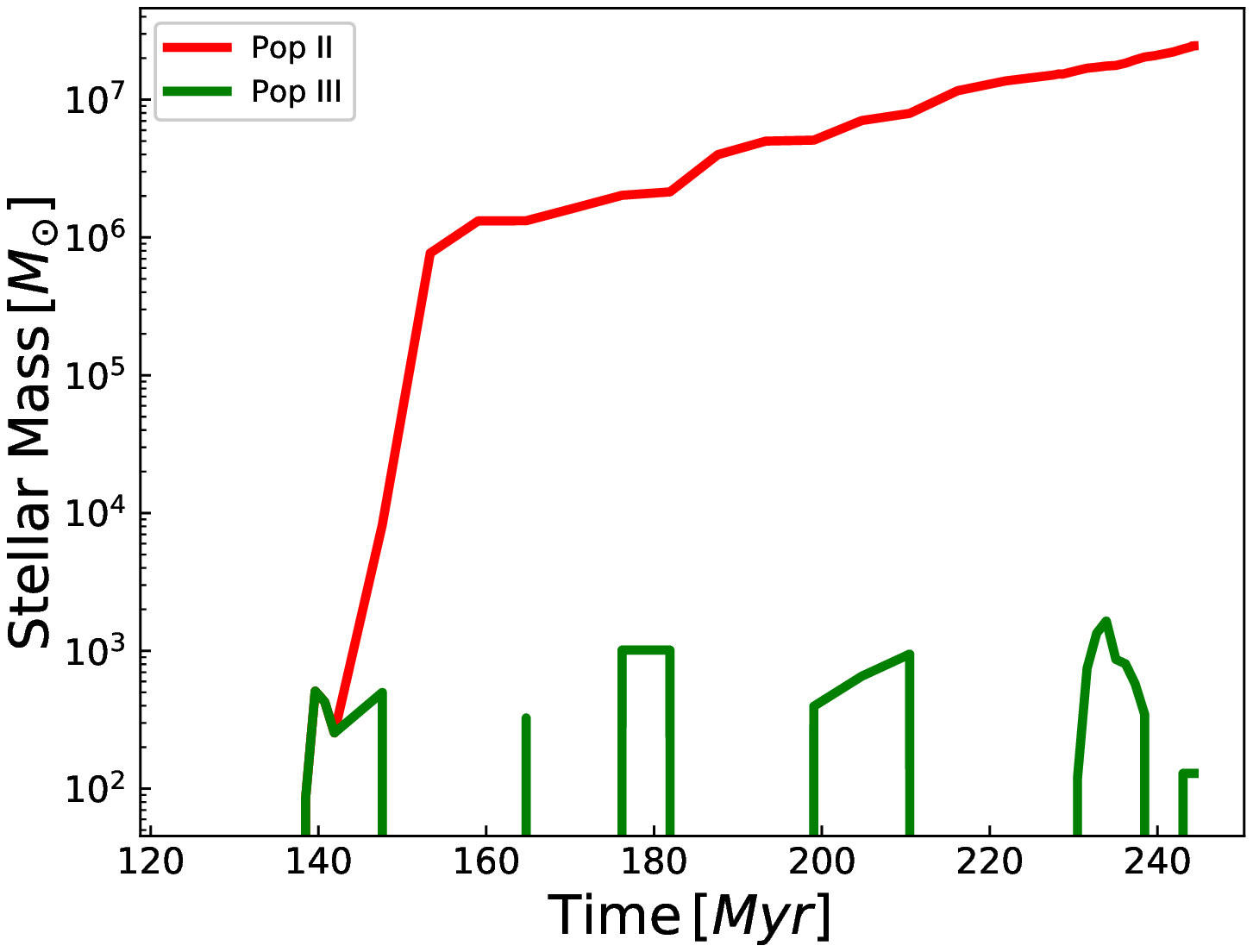}
\end{minipage} &
\begin{minipage}{6cm}
\hspace{1.8cm}
\includegraphics[scale=0.54]{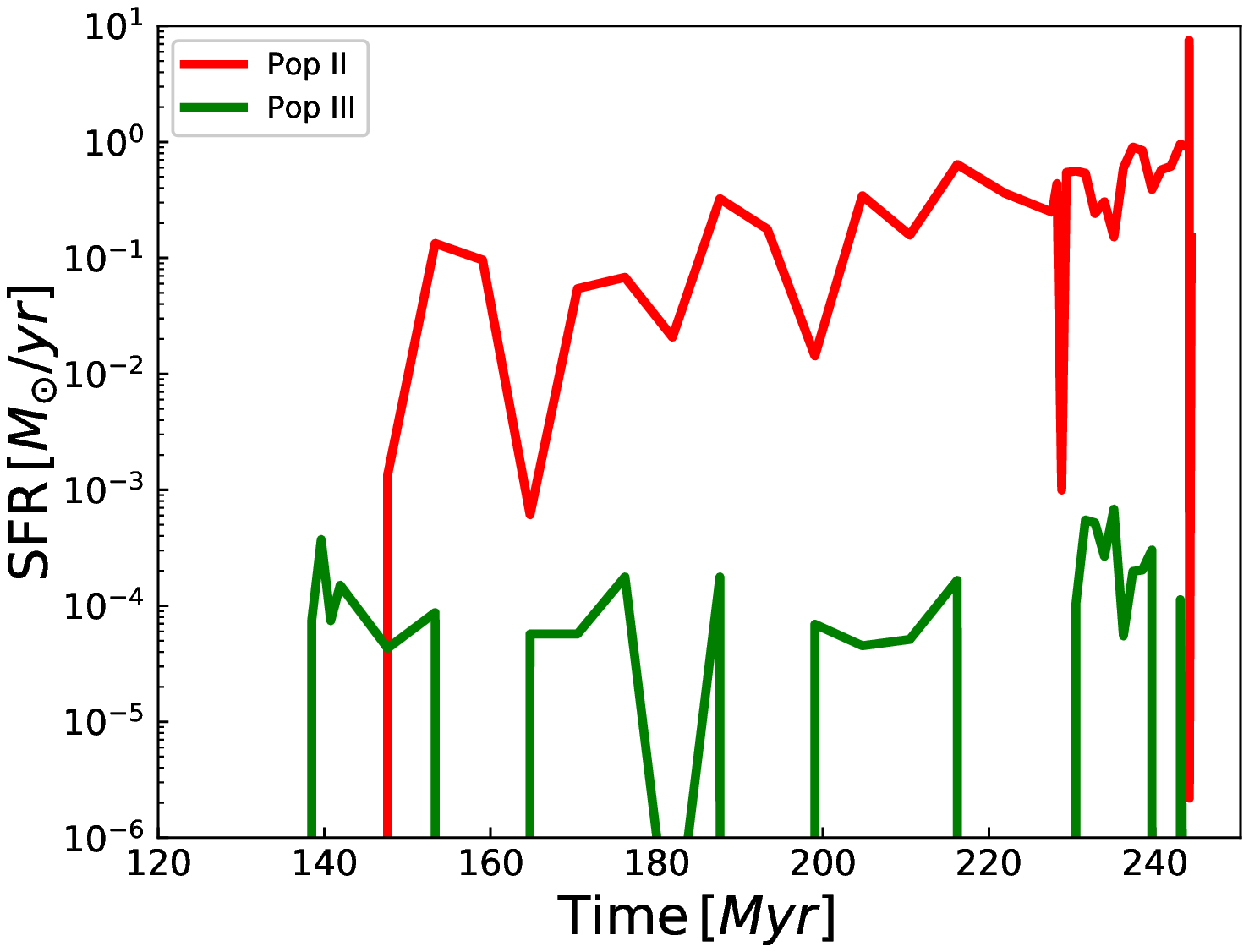}
\end{minipage} \\ 
\begin{minipage}{6cm}
\vspace{-0.2cm}
\includegraphics[scale=0.54]{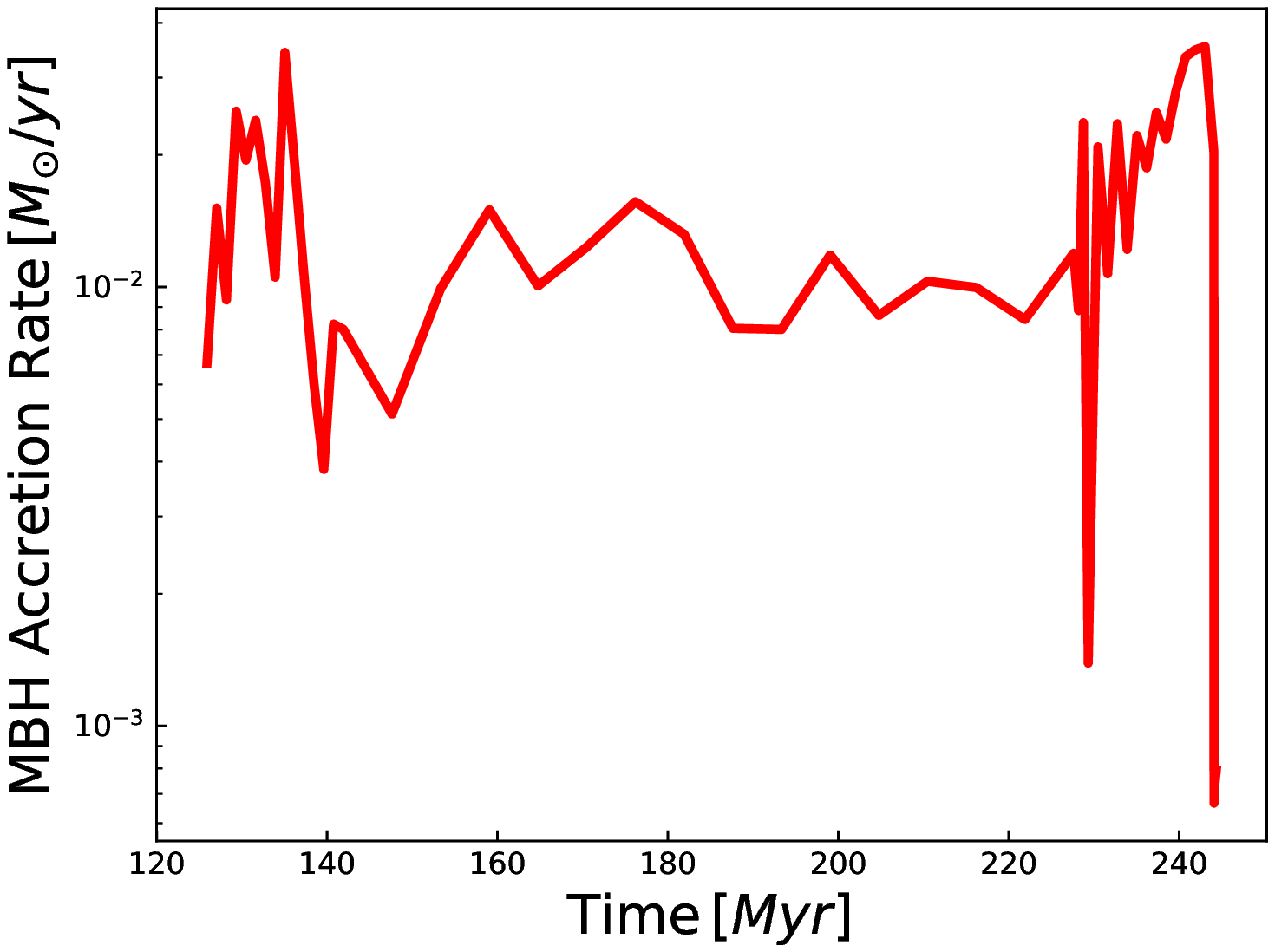}
\end{minipage} &
\begin{minipage}{6cm}
\hspace{1.8cm}
\includegraphics[scale=0.54]{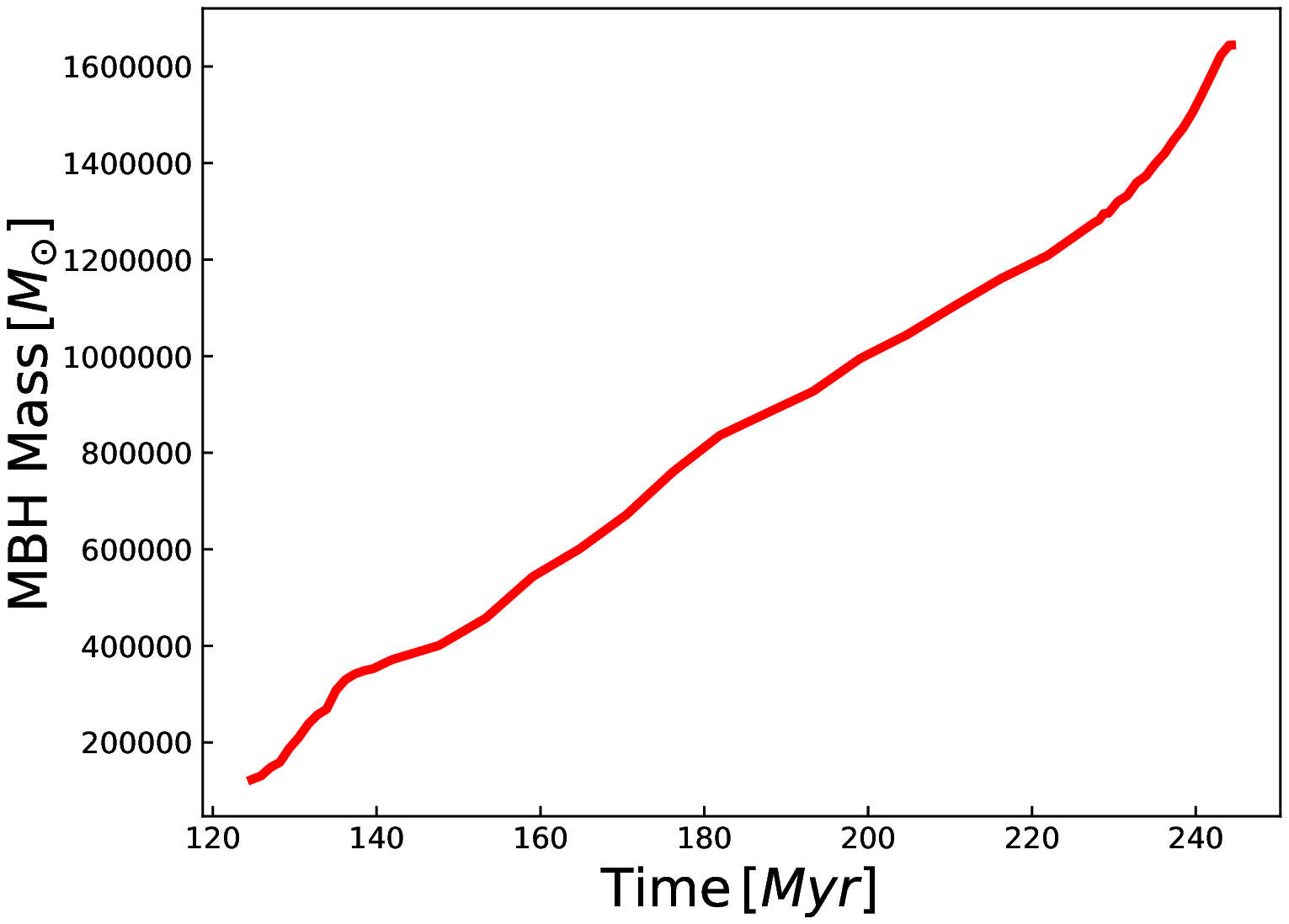}
\end{minipage} \\
\begin{minipage}{6cm}
\vspace{-0.2cm}
\includegraphics[scale=0.54]{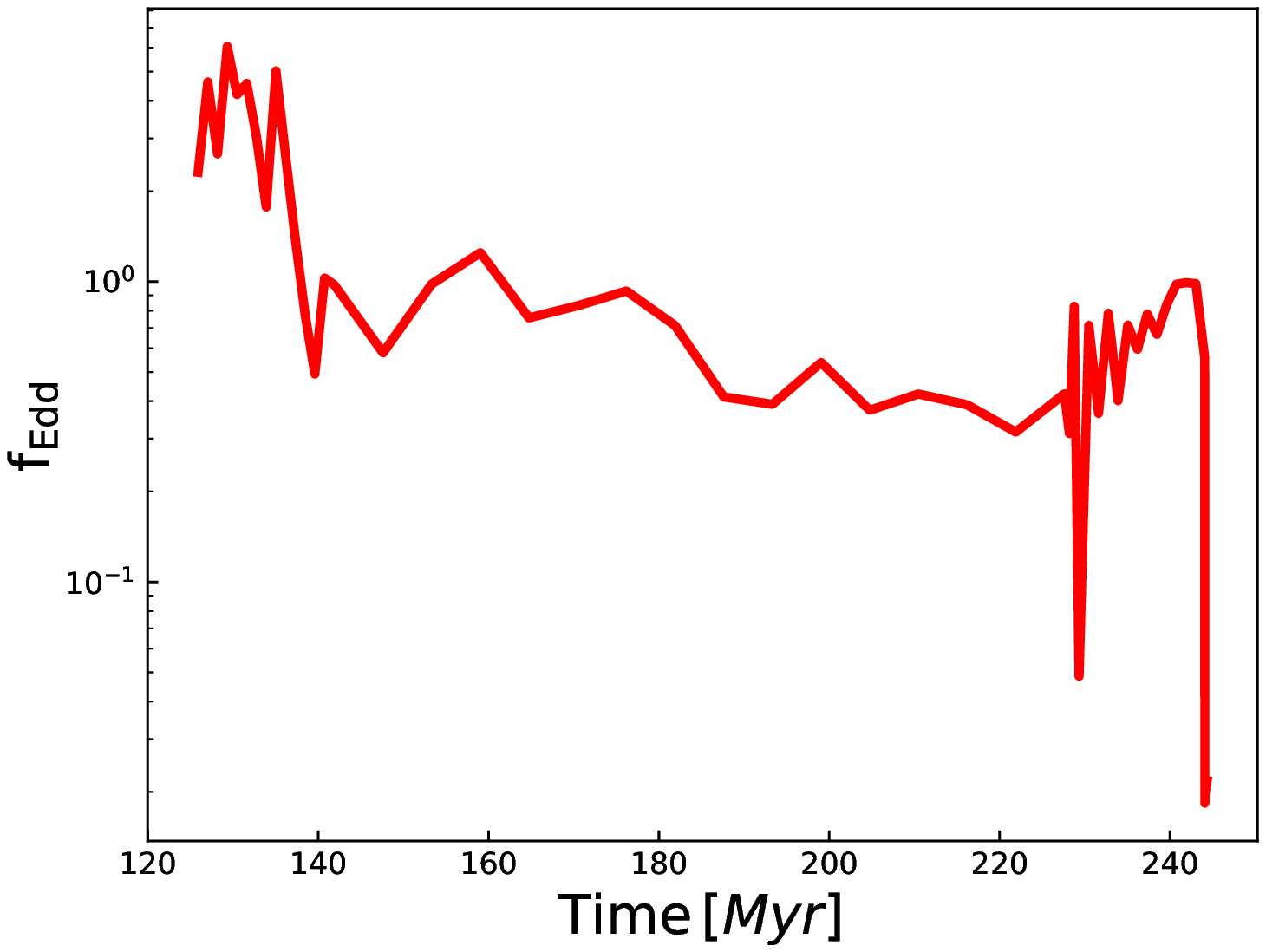}
\end{minipage} &
\begin{minipage}{6cm}
\hspace{2.2cm}
\includegraphics[scale=0.54]{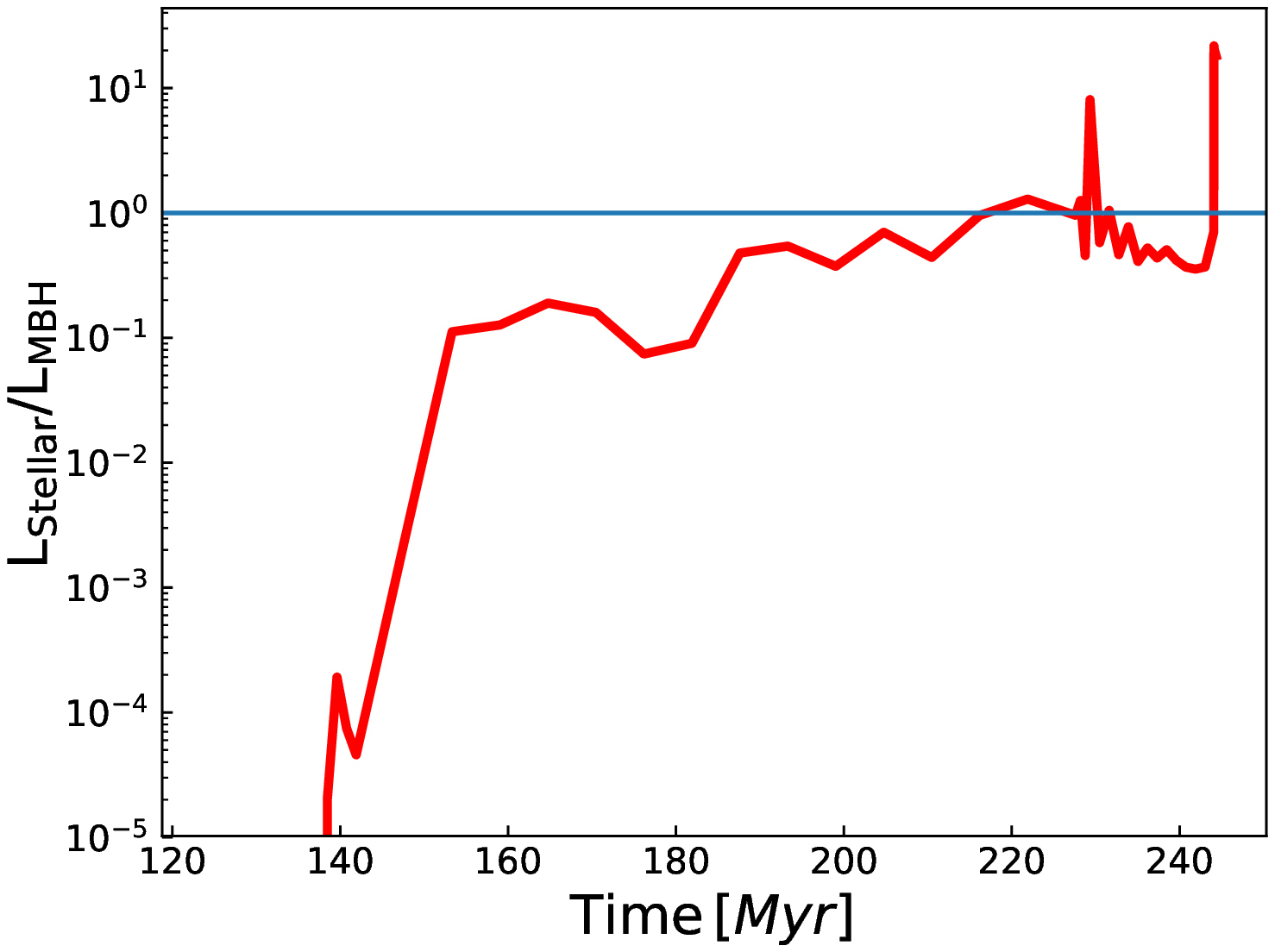}
\end{minipage} 
\end{tabular}
\caption{Stellar mass  and SFR vs. time (top panels), mass accretion rate onto MBH and mass of MBH vs. time (middle panels), the Eddington ratio and  the ratio of  Stellar to  MBH luminosity  vs. time (bottom panels). Time is shown in Myr after the Big Bang. }
\label{fig}
\end{figure*}

\begin{figure*} 
\hspace{-6.0cm}
\centering
\begin{tabular}{c c}
\begin{minipage}{6cm}
\vspace{-0.2cm}
\includegraphics[scale=0.5]{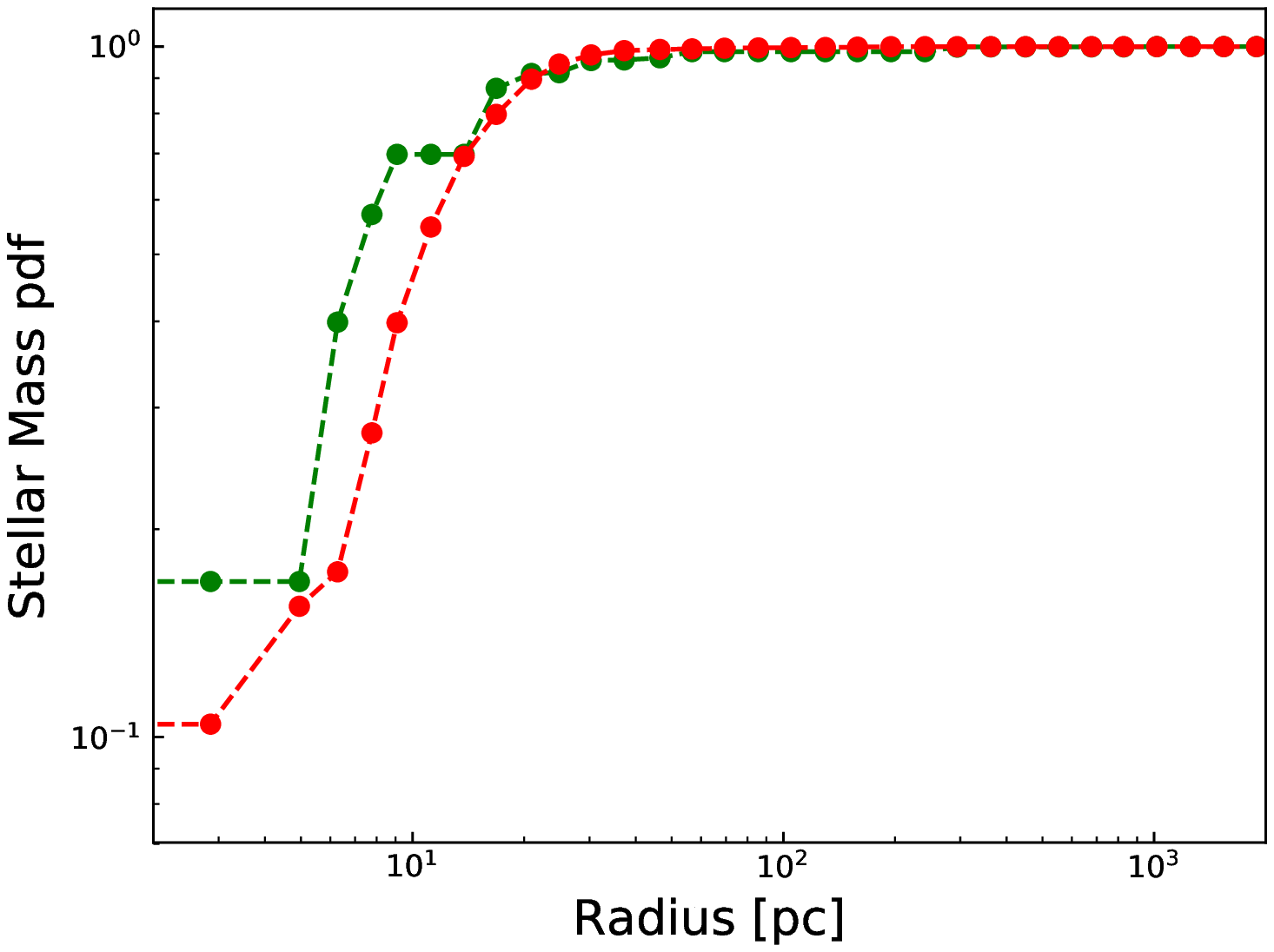}
\end{minipage} &
\begin{minipage}{6cm}
\hspace{1.8cm}
\includegraphics[scale=0.5]{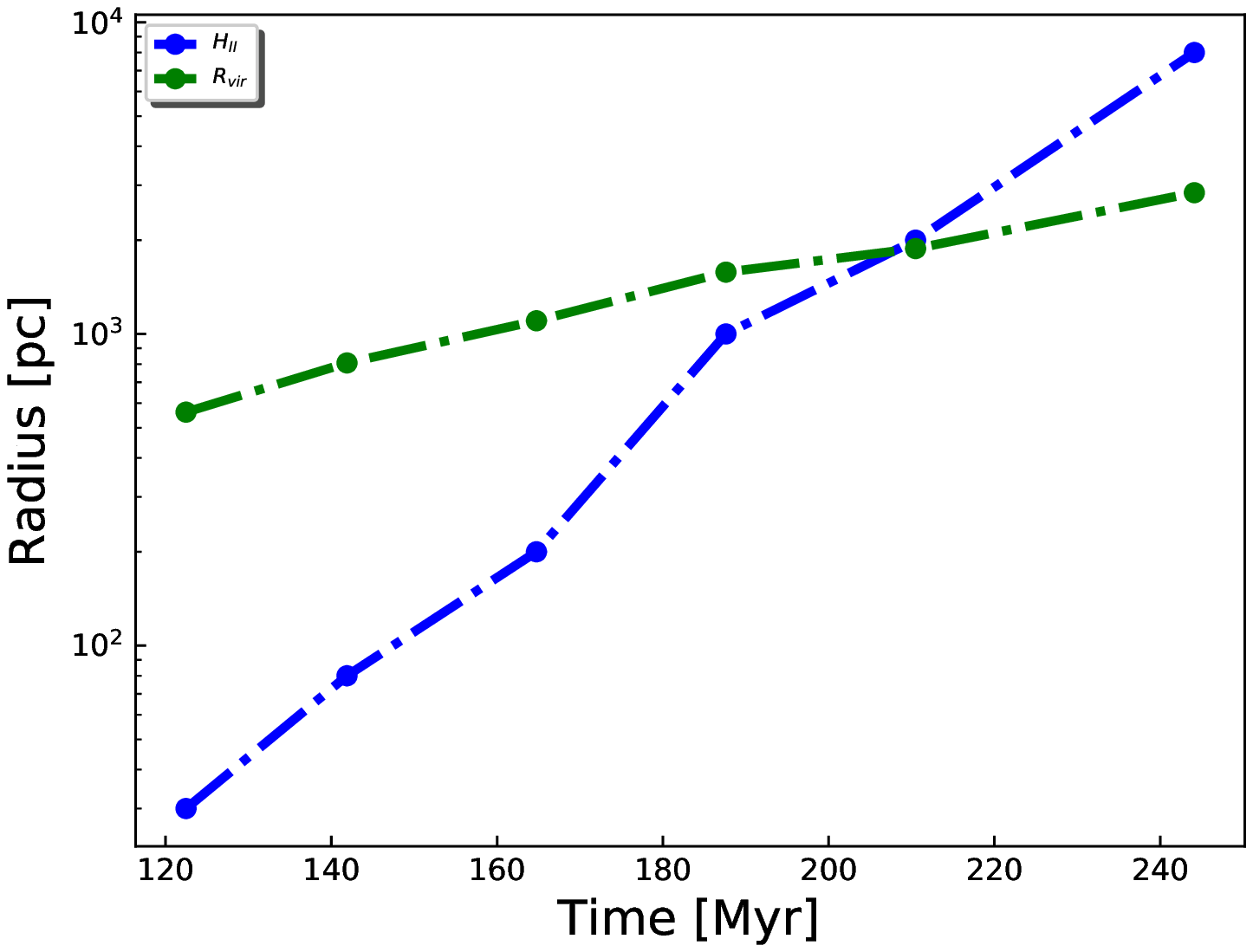}
\end{minipage} \\
\begin{minipage}{6cm}
\vspace{-0.2cm}
\hspace{3cm}
\includegraphics[scale=0.54]{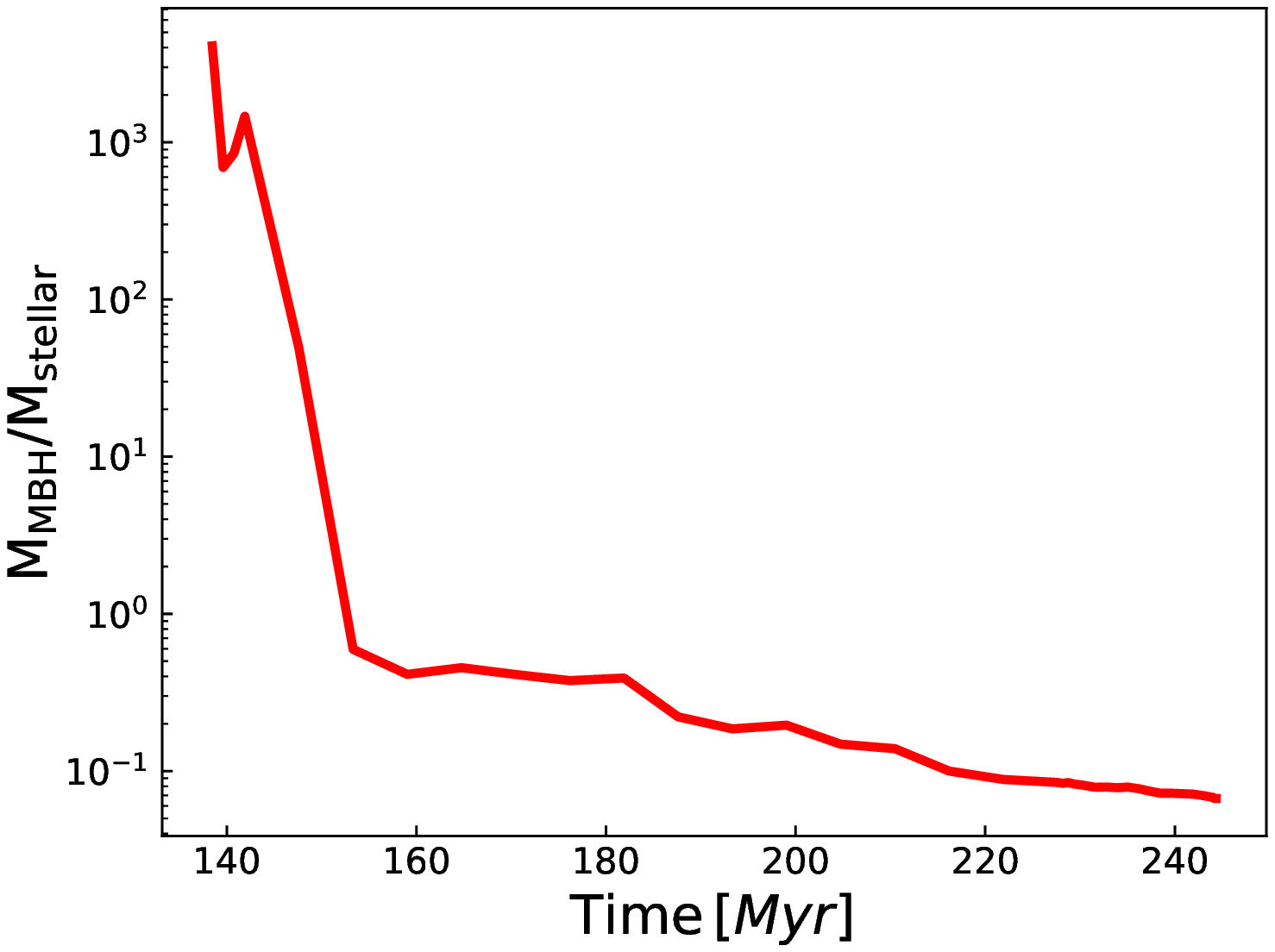}
\end{minipage}
\end{tabular}
\caption{The normalized cumulative stellar mass distribution function of stellar birthplaces is  shown in the upper left panel and the right panel shows the evolution of $\rm H_{II}$  against time. The bottom panel shows the ratio of MBH mass  to  total stellar mass  against time. Time is shown in Myr after the Big Bang. }
\label{fig1}
\end{figure*}

 On the  contrary,  metal rich gas from Pop III SNe gets cooled over timescales of a few Myr as explosions occur in the dense gas which yield continuous  Pop II SF in the main halo.  This results in Pop II stellar mass of $\rm 2 \times 10^7~M_{\odot}$ in 100 Myr as shown in Fig. \ref{fig}. The average Pop III SFR during bursts is  $\rm \sim 10^{-4}~M_{\odot}/yr$ and  the Pop II SFR is $\rm \sim 0.2~M_{\odot}/yr$ and  increases up to $\rm \sim 1~M_{\odot}/yr$ over the last 40 Myr. The overall increase in SFR for Pop II stars is related to the halo growth which has increased by two orders of magnitude in 100 Myr.  Overall the halo grows via accretion  and minor mergers of pristine gas. Pop III SF is bursty, they  produce metals through SNe  which cool down over a time scale of a few Myr and lead to  insitu Pop II star formation due to the efficient mixing. 
 To quantify the spatial distribution of stars in the galaxy, we have computed the cumulative stellar mass probability distribution function for the birth places of Pop III and Pop II stars shown in Fig. \ref{fig1}. This suggests that most of the SF occurs in the central 100 pc of the galaxy and  Pop III stars are more centrally concentrated compared to Pop II stars. The  SF in combination with the accreting MBH continously heat  the inter-stellar gas, consequently some of these photons may leave the galaxy and ionize the gas in the surrounding medium. Since our simulation employs radiative transfer on-the-fly from all stars and MBH, contrary to commonly used thermal feedback, it provides better estimates of  ionized gas. The  galaxy wide HII region is anisotropic, intermittent and ionizing radiation preferentially  leaks into the low density medium. We also estimate the evolution of  region with at least 10 \% of ionized gas in the enclosed radius and compare it to the viral radial of halo. These estimates are shown in Fig \ref{fig1} and suggest that  region extends beyond the virial radius of halo at $\sim$ 210 Myr.

\begin{figure*} 
\hspace{-6.0cm}
\centering
\begin{tabular}{c c}
\begin{minipage}{6cm}
\vspace{-0.2cm}
\includegraphics[scale=0.5]{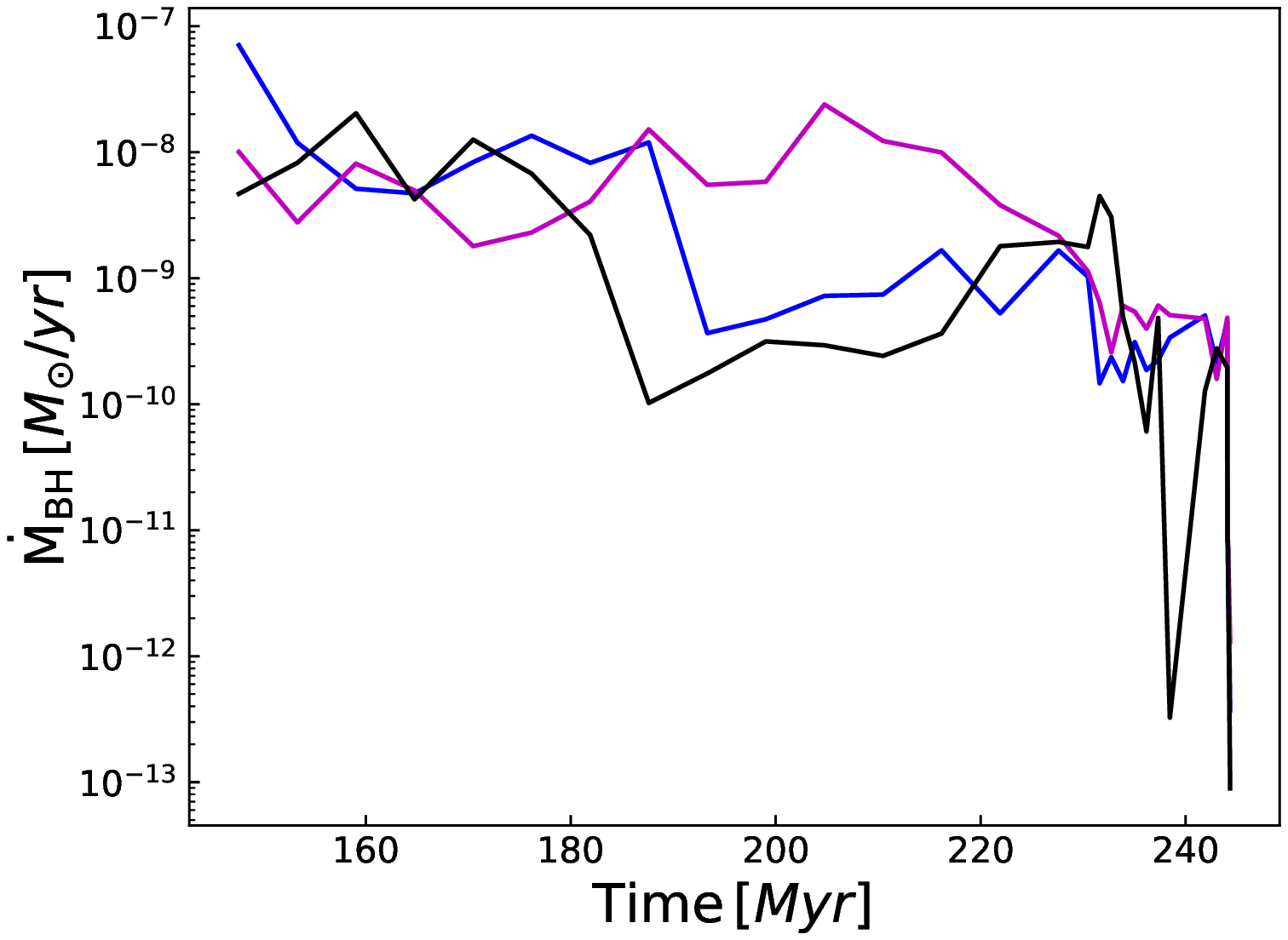}
\end{minipage} &
\begin{minipage}{6cm}
\hspace{1.8cm}
\includegraphics[scale=0.5]{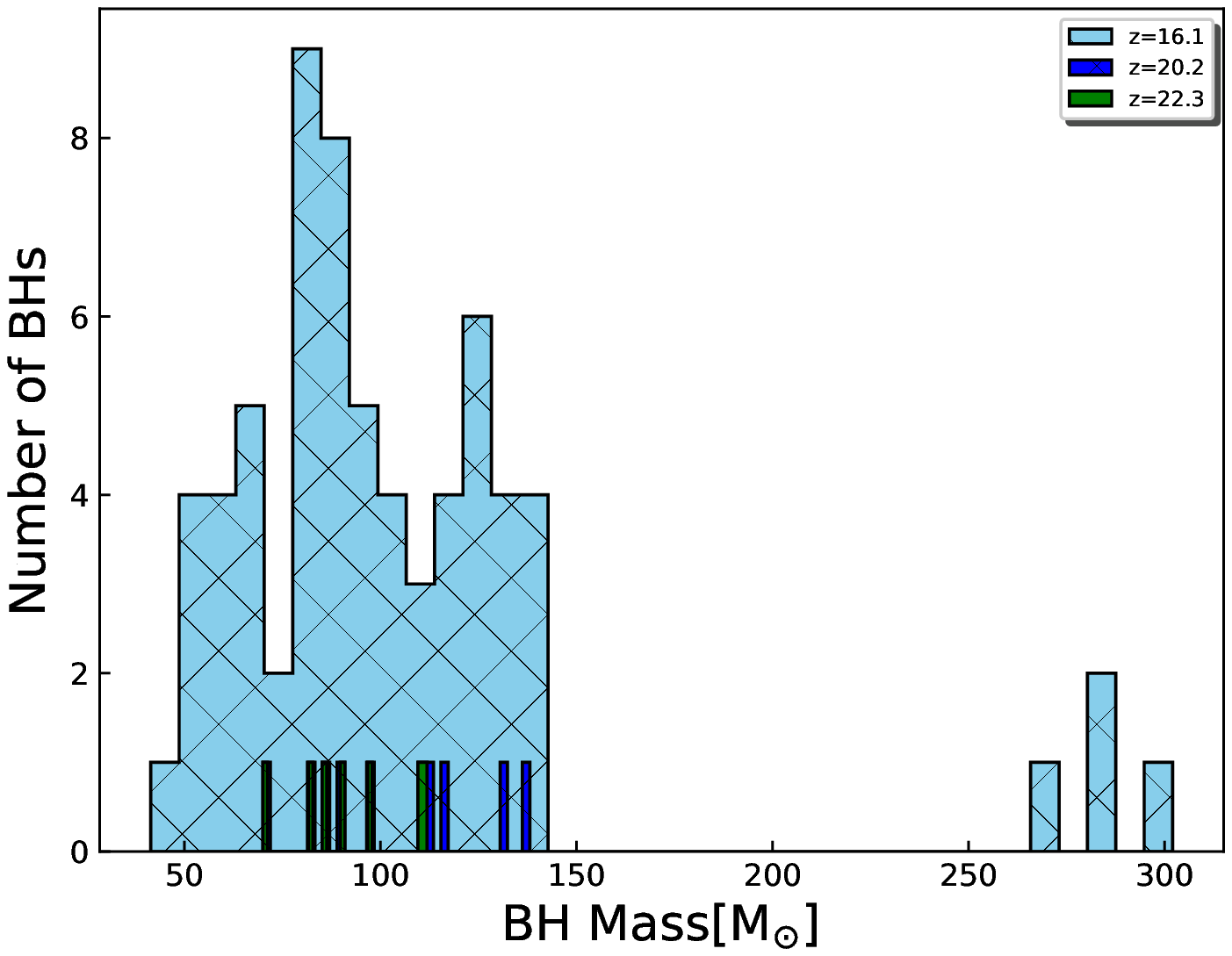}
\end{minipage}
\end{tabular}
\caption{The accretion rates onto three representative stellar mass BHs  (left panel) and  mass distribution of Pop III remanant BHs at  $z=22.3$, $z=20.2$ and $z= 16.12$.}
\label{fig2}
\end{figure*}

\begin{figure*} 
\begin{center}
\includegraphics[scale=1.0]{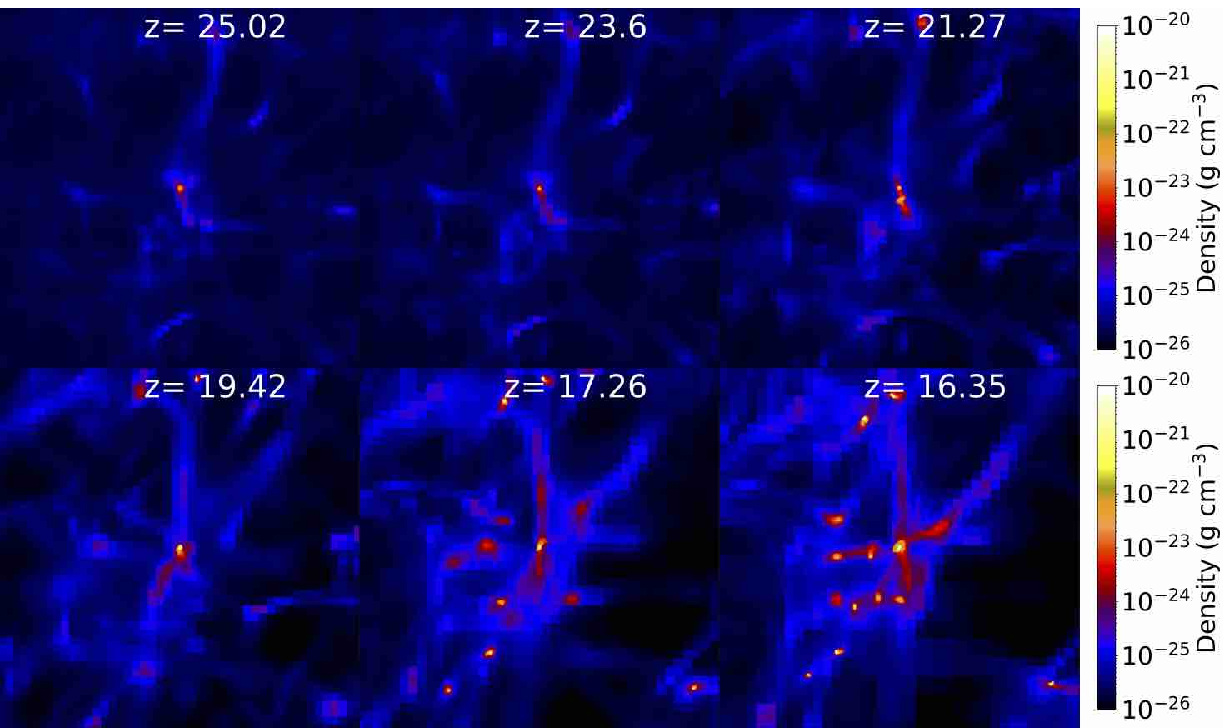}
\end{center}
\caption{The average density along the line of sight is shown for 10 kpc region centered at MBH for different redshifts.}
\label{fig3}
\end{figure*}

\subsection{MBH growth}

We seed  a MBH  of $\rm 10^5~M_{\odot}$ at the center of the first atomic cooling halo at $z \sim 26$. The mass accretion  onto the MBH kicks off at  a few times $\rm 10^{-3} ~M_{\odot}/yr$,  X-rays from the MBH  heat the gas in its proximity and push it away in a wind with typical velocities of 40 km/s. However, mass accretion onto the MBH continues due to gas infall. This drives  intermittent accretion for the first 15 Myr in the absence of  stellar feedback with the peak mass accretion rate of $\rm 0.03 ~M_{\odot}/yr$  see Fig. \ref{fig}. This corresponds to the averaged accretion rate of  4 times  Eddington accretion during first 15 Myr.  Although accretion rates are numerically restricted to the Eddington limit in our simulation, this phase of super-Eddington accretion is due to the immediate merging of star particles forming in the host cell of the MBH.  The density around the MBH  ranges  from $\rm1- 10^4 ~cm^{-3}$,  see Fig. \ref{fig3}. The first Pop III SN occurs at 140 Myr in the vicinity of the MBH, followed by a few more SNe which  heat the gas and create an outflow resulting in about an order of magnitude decrease in the mass accretion rate.  Since the MBH is embedded in dense accretion flows, the metal rich gas gets cooled in the aftermath of SNe and starts to feed the MBH again increasing the accretion rate  back to $\rm 0.01 ~M_{\odot}/yr$.  The fluctuations in  the mass accretion rate are correlated with the bursty mode of Pop III SF.  Pop III stars go off  SNe,  heat the gas, generate outflows and evacuate the gas  from the vicinity of the MBH. Such clumpy accretion continues throughout the simulation.  The metal rich gas cools on short timescales and falls back onto MBH. The upward trend in the mass accretion rate during the last 20 Myr is a consequence of  multiple dense clumps merging with the main halo which boost accretion onto the MBH.   They trigger  starbursts at  229 and 242 Myr, the radiative feedback from young stars generates  outflows a which evacuate the gas from the surrounding of the MBH and  the accretion rates briefly drops by almost an order of magnitude.  Overall, the mass accretion onto the MBH is highly intermittent and averages to $\rm 0.015 ~M_{\odot}/yr$ across the 120 Myr of our simulation.

The average mass accretion rate onto the MBH is four times the Eddington during  the first  15 Myr in the absence of stellar feedback and averages to $\rm $50 \% Eddington over the course of 120 Myr, see Fig. \ref{fig}.  Consequently the mass of the MBH grows at a faster rate in the absence of stellar feedback, slows down  during outflows produced from Pop III SNe  and then almost linearly increases with time. The small jumps in the MBH mass are due to the clumpy accretion and about 400,000 $\rm M_{\odot}$ is accreted onto the MBH during the last 20 Myr resulting in an average accretion rate of  $\rm 0.02 ~M_{\odot}/yr$. This increase is attributed to merging of dense  clumps at these times and a deeper potential well  retaining more gas because of the increase in halo mass. In total, $\rm 1.5 \times 10^6~M_{\odot}$ has been accreted onto the MBH during the course of 120 Myr with the mean accretion rate of  $\rm 0.01 ~M_{\odot}/yr$.  To quantify the contribution of  stellar vs. MBH radiative feedback, we  estimated the ratio of the stellar luminosity  from young stars to the MBH accretion luminosity which is shown in the bottom right panel of Fig. \ref{fig}.  Our results show that the X-ray luminosity from MBH dominates over stellar radiative feedback  most of the times and the MBH  growth is regulated by SNe feedback in tandem with X-ray heating. However, during the last 20 Myr the  contribution from stellar radiative feedback becomes comparable to the energy released by  X-ray feedback from the MBH and even regulates the mass accretion onto MBH during starburst phases. The latter is evident from the fact that peaks in the stellar to MBH luminosity ratio  are directly correlated with  drops in the MBH accretion rates. In summary, there are two main processes in play which feed the MBH, pristine gas accretion from the cosmic web and the recycled metal enriched gas from SNe.  The metal rich gas cools over short timescale and mainly feeds the MBH. This is evident from the fact that gas accretion from the cosmic web is episodic, occurs  after every 20 Myr and  in the mean time accretion occurs onto the MBH  via metal enriched gas. 

\subsection{Galaxy and MBH coevolution}

Our simulation is one of the first  to explore the co-evolution of a host galaxy and massive black hole forming at $z=26$ down to $z=16$ by self-consistently modeling the star formation, stellar feedback along with X-ray feedback  from the MBH.  Our simulated galaxy has stellar mass of  $\rm 2 \times 10^7~M_{\odot}$ at $z \sim16 $ and  MBH of $\rm 1.6 \times 10^{6}~M_{\odot}$.  To understand this co-evolution, we have computed the MBH  to total stellar mass ratio shown  in Fig. \ref{fig1}. We find that MBH mass dominates the stellar mass for  first 30 Myr after its birth,  the MBH to total stellar mass ratio sharply declines down to 0.3 as Pop III star go off SNe, enrich the medium with metals and boost Pop II star formation.  This ratio  gradually keeps dropping as both the host galaxy and the central MBH grow hand in hand.  The  MBH to total stellar  mass ratio is  $\rm 5 \times 10^{-2}~M_{\odot}$ at the end of our simulation compared to local BH-stellar mass relation: $\rm M_{BH}/M_{stellar} \sim  2.5 \times 10^{-4}$  \citep{Reines2015}. This provides a unique correlation between the stellar and the MBH mass in DCBH hosting galaxies at $z=16$.

Pop III stars with masses between $\rm 40-140~M_{\odot}$ and larger than 260 $\rm M_{\odot}$ collapse into BH. We have plotted the mass distribution of BHs resulting from the collapse of Pop III stars in Fig \ref{fig2} at three different times. In total, we have 67 BHs in our galaxy at $z=16$, the typical mass accretion rates onto the BHs are a few times $\rm 10^{-9}~M_{\odot}$ and drop further by an order of magnitude at the end of simulation, see left panel of Fig \ref{fig2}.  The prevalence of such low accretion rates suggests stunted growth of stellar mass  BHs  as none of them is able to grow efficiently for 120 Myr. They are formed in HII regions created by massive Pop III stars and X-ray feedback from them further evacuates  gas from their vicinity which hampers their growth.  This is consistent with previous studies exploring the growth of stellar mass BHs \citep{Johnson2007,Smith18}.

\section{Discussion and conclusions}

We have conducted a  cosmological hydrodynamical simulation coupled with the 3D radiative transfer module MORAY to  model X-ray feedback from MBH and UV radiative feedback from Pop III and Pop II stars. In our simulation we self-consistently include the formation of Pop III and Pop II stars along with their chemical, mechanical and radiative feedback in the host galaxy. We insert a MBH of $\rm 10^5~M_{\odot}$ in an atomic cooling halo at  $z=26$  and follow its growth along with the assembly of the host galaxy  down to $z=16$ for 120 Myr in a halo of $\rm 2 \times 10^9~M_{\odot}$. Our simulation has a physical resolution of a few parsec and DM resolution of $\rm 10^5~M_{\odot}$. We find that X-rays  from an accreting MBH play a dual role,  they delay Pop III SF for the first 12 Myr and in meantime catalyze $\rm H_2$ formation which later induces SF. The MBH accretes at 300 \% Eddington in the absence of stellar feedback and its growth is mainly regulated by SNe in tandem with X-ray heating.  Pop III SNe  enrich the galaxy with metals and consequently the Pop II stellar mass sharply increases to $\rm \sim 10^6~M_{\odot}$ within a few Myr after their formation.  Pop II SF continues in the host galaxy and the  stellar mass  at $z=16$ is $\rm 2 \times 10^7~M_{\odot}$. The Pop III star formation is episodic and bursty while Pop II stars form continuously. Overall, the MBH accretes $\rm 1.5 \times 10^6~M_{\odot}$ in 120 Myr with an average mass accretion rate of $\rm 0.01~M_{\odot}/yr$.  In our simulation stellar mass BHs do not grow as mass accretion rates on them range from a few times $\rm 10^{-10}-10^{-9}~M_{\odot}$.  The stunted growth of Pop III remnant BHs  is due to their formation in HII regions created by massive stars and later X-rays further evacuate the gas from their vicinity.

Our results  show that  X-rays  from MBH induce SF by boosting molecular hydrogen and are in agreement with results from  \cite{Aykutalp19}. We have studied here the early phase of  MBH growth from $z=26~to~z=16$ which has remained unexplored in previous work.  In comparison with MBH growth in halos of $\rm \sim10^8~M_{\odot}$ \citep{Aykutalp14}, we find that accretion rates are about two orders of magnitude higher due to the deeper halo potential well.  \cite{Latif18} explored the growth in haloes of $\rm 3 \times 10^{10}~M_{\odot}$ at $z=7.5$. Stunted MBH growth in their work  could be due to the delayed SF that lead to strong starbursts  which generated galaxy wide outflows, evacuated the gas from the MBH vicinity and also they ignored metal cooling from Pop II SNe. The mass accretion rate onto MBH in  \cite{Smidt17}  at $z=14$ is $\rm \sim 0.2~ M_{\odot}/yr$, an order of magnitude higher than in our  work at $z=16$. The differences from the former study could arise due to the modeling of Pop III SNe in our simulation as they are more energetic and can remove gas more efficiently from haloes and is ignored in their work. Large scale cosmological simulation on other hand  are unable to capture this early phase of growth due to their limited resolution \citep[e.g.][]{Booth209,Dubois15,Sijacki15,DiMatt17}.

We have ignored here mechanical feedback from MBH  arising from jets. Recently, this was investigated by \cite{Regan19}, they found that  the impact of bipolar jets is localized to sub-pc scales and  does not break out from hosting atomic cooling haloes. Therefore we expect that the impact of mechanical feedback would be negligible on our findings. We have employed here the Bondi-Hoyle recipe to model the accretion onto MBH which does not take into account angular momentum \citep{Debuhr10}.  In future a detailed comparison between the Bondi-Hoyle and the alpha-disk models is required to assess their impact on MBH growth.  

We have assumed here that X-ray luminosity from a MBH arises at 2 keV.  This choice corresponds to the  characteristic temperature of the quasar spectral energy distribution (SED),  estimated by equating Compton heating and Compton cooling by the given SED \citep{Sazonov04}. Therefore it can be considered as the temperature of a hot plasma in the vicinity of the MBH. We  do not model full SED from an accreting MBH which are expected to have also emission in UV and infrared \citep{Yue13,Pacucci15}. Some of low energy photons are expected to get attenuated in the MBH vicinity and therefore we expect their impact to be moderate. We also assumed here that  all radiation  from MBH is emitted at 2 keV and the higher energy X-rays will escape the  halo due to their long mean free paths.  In fact, about 45 \% of the total bolometric luminosity  \citep{Yue13} is emitted in X-rays and therefore, we may overestimate the impact of X-ray by a factor of two. In future  we plan to employ multi-color template  to model SED from an accreting MBH.

Overall the MBH in our simulation has grown mainly through  gas accretion and we expect this mode of growth to be common at high redshift. The upcoming X-ray observatories such as ATHENA and Lynx may help in better understanding the growth of MBHs by observing a few hundred  low luminosity AGN upto $z=20$ \citep{Pacucci20}. Also,  such MBHs can be observed  in infrared with the James Webb Space Telescope at $z<20$ and even also be detected in radio with next-generation Very Large Array and the Square Kilometer Array \citep{Whalen20a,Whalen20b}.

\section*{Acknowledgements}

MAL thanks the UAEU for funding via startup grant No. 31S372 and UPAR grant No. 31S390. We thank the Dominik Schleicher for helpful discussions.



\section{Data Availability Statement}
The data underlying this article will be shared on reasonable request to the corresponding author.

\bibliography{smbhs.bib}





\bsp	
\label{lastpage}
\end{document}